# The Evolution of Assistive Technology: A Literature Review of Technology Developments and Applications

Matteo Zallio[1] and Takumi Ohashi[2]

[1] University of Cambridge. Department of Engineering, Engineering Design Centre, United Kingdom
mz461@cam.ac.uk

[2] Tokyo Institute of Technology. 2-12-1 Ookayama, Meguro-ku, Tokyo 152-8550, Japan
ohashi.t.af@m.titech.ac.jp

**Abstract:** The term "Assistive Technology" (AT) has evolved over the years and identifies equipment or product systems, whether acquired, modified, or customized, that are used to increase, maintain, or improve functional capabilities of individuals with disabilities. Considering the advances that have been made, what trends can be identified to provide evidence of the evolution of AT as devices that foster accessibility and empower users with different abilities? Through a systematic literature review we identify research items that offer evidence of the evolution of the meaning, purpose, and applications of AT throughout the history. This paper provides evidence that AT evolved from products to improve functional capabilities of individuals with disabilities toward enabling technologies that facilitate tasks for people with different needs, abilities, gender, age, and culture. This evolution will lead to a positive demystification of the meaning and applications of AT toward broad usage acceptance among mainstream users.

## 1    Introduction

The term "Assistive Technology" (AT) has been widely used over the years and within various domains. One of the first official definitions of AT was included in the Technology-Related Assistance for Individuals with Disabilities Act, which was first passed in 1988, reauthorized in 1994, and reproposed in 1998 [1]. According to the act, the term AT identifies any item, piece of equipment, or product system, whether acquired commercially, modified, or customized, that is used to increase, maintain, or improve functional capabilities of individuals with disabilities [2]. Additionally, there is an emphasis on the term AT service, which refers to any service that directly assists an individual with a disability in the selection, acquisition, or use of an AT device [3].
As technology has become one of the primary engines for economic activity, education, and innovation, the substantial progress made in the development of AT has significantly benefited individuals with disabilities of all ages. Such devices and adaptations have increased the involvement of such individuals in society, and have reduced expenditures associated with education, rehabilitation and training, employment, residential living, and independent living [4] . With the implementation of the 1994 act, followed by the 1998 act, further concepts were added to the definition of AT. The revised definition from the 1998 act provided a broader concept of AT, which included a wider range of applications and potential users.

AT was not merely used to augment capabilities for individuals with disabilities, but also could support other persons, including family members and caregivers to perform daily contextual tasks associated with instrumental activities of daily living including education, work, social and cultural life [3].

Six years later, the Assistive Technology Act of 2004 amended the previous act of 1998 by embracing the definition of AT as a service consisting of expanding the availability of access to technology, including electronic and information technology to individuals with disabilities [5].

Reading through these definitions, the broader meaning and scope of AT is clear: to deliver a product and a service to support people with different capabilities and help to improve their overall life conditions.

Work resulting from the World Health Organization's Global Collaboration on Assistive Technology (GATE) and the Global Research, Innovation and Education on Assistive Technology (GREAT) aimed to maximize the impact of AT in enabling participation as a holistic understanding of the value and meaning of AT for the individual, focusing on the person first, then considering its conditions and context.

The advancement of the scope of AT over the past three decades highlights a significant evolution of the definitions and applications of products and services in the AT domain by embracing a range of people with different capabilities who could benefit from them [6,7].

In recent years, more technologies have been developed and certain ones initially developed for specific users including individuals with mild or moderate impairments, or older adults, evolved as augmentative products for mainstream users [8].

By looking at this evolution of the definitions and applications of AT, a question arises. What trends and applications of AT can be identified across the scientific literature to provide evidence of the growth of AT as devices that foster accessibility and empower users with different abilities?

In this paper we aim to identify the trends in the evolution of the meaning, purpose, and applications of AT with the aim of defining perspectives of AT that consider current trends. Through a systematic literature review we aim to develop evidence-based knowledge of the past and current views on AT and identify a scenario that will support AT to become an accepted, mainstream term to allow for a decreased stigmatization among individuals that use a variety of assistive devices.

## 2    State of the art: history and context

AT was recently defined by ISO standard 9999:2016, Assistive products for persons with disability [9] as a group of technologies including devices, equipment, instruments and software, especially produced or generally available, used by or for persons with disabilities for different purposes.

Although this recent definition seems to be broadest to date, there are different examples, dating back to centuries ago, of early developments of AT that refer to such a definition and are related to different applications. From simple technological innovations such as eyeglasses, developed in Italy around 1200 [10], to wheelchairs first developed in China in the 5th century [11], a large variety of items can be classified under the term AT.

To help identify some examples of AT as products or equipment to support people with disabilities, it is important to refer to the International Classification of Functioning (ICF), which defines different capabilities, including vision, hearing and speech communication, mobility, and cognition and learning skills [12].

Some instances of AT to support vision abilities could refer to products that augment visual capability, allowing for better performance of certain Activities of Daily Living (ADLs). Braille displays allow people who have moderate and severe visual impairments to access spaces and understand how to use certain objects [13].

Magnifiers offer great support for people who have minimal or mild visual impairments to see images, read text, and identify meaningful information more clearly [14].

Text-to-speech systems and screen-reading software are supporting technologies that aim to support a wide variety of people with minimal, moderate, or severe visual impairments [15].

Regarding hearing and speech communication, there have been a number of devices developed specifically for deaf or hard-of-hearing people, including AT used to support ADLs [16]. Some of the most common examples are personal amplification systems for older adults or people with hearing loss.

In terms of more mainstream products that have been accepted and appreciated by users, other systems for close captioning for online meetings, TV shows, and phone calls can be identified. Additionally, visual feedback provided, for instance, by flashing lights or a visual alert system for a doorbell or phone are other interesting examples of technologies initially developed for people with hearing or speech communication challenges.

Mobility, comprising upper and lower body mobility, includes different technologies developed throughout the centuries [17]. It is possible to recall several devices to support people with moderate or severe mobility impairments that use power chairs, wheelchairs, and walkers. However, people who experience a temporary disability or a mild disability could use a walking stick or crutches. Three- or four-wheel scooters, which have become more widely accepted and used in different countries, can allow people with different levels of impairments to visit the neighborhood and surrounding areas.

In recent years, several technologies to support cognitive and learning skills have been developed to allow people with different levels of skills to perform daily routines at home or in the community [18]. Products that help people to enhance attention, memory, and organization can be summarized as memory aids and reminders, text-to-speech systems for improved learning (not related to vision), and specialized apps to create text from audio files. These are just some examples of AT, developed to address different challenges as described by the ICF, and it is possible to identify a pattern that shows different levels of specifications that AT has used to solve precise challenges that different people experience. In this non-exhaustive list, several products were specifically developed for people with severe impairments and more recently it appears that the same products have become more mainstream and appreciated by a wider audience. This observation of trends generates a hypothesis to be verified.

## 3  Research Methodology

### 3.1 Data Collection

To clearly identify and provide evidence-based support for the hypothesis we propose a systematic literature review of scientific publications to understand the type of technology developed, its application in terms of users and capabilities, its evolution and adoption over time, and what trends have emerged across a consistent number of research projects. The systematic literature review was conducted based on the PRISMA model [19] and was run on October 22, 2021, using the Web of Science Core Collection database. The literature review search focused only on English-language articles and followed different steps. The first phase was articulated with the following search query: [(TI = (assistive technology)) AND TI = (vision impair* OR near-blind OR partially sighted OR visually challenged OR hearing impair* OR hard of hearing OR deaf OR mobility impair* OR cognitive impair* OR cognitive disorder)]. This search resulted in a collection of 1635 papers.

The literature review included articles from the engineering-related field; thus, the first narrowing down of papers consisted of selecting the following areas of engineering in the Web of Science search engine, resulting in 1197 papers.

*Rehabilitation*, *Engineering Electrical Electronic*, *Computer Science Cybernetics*, *Computer Science Information Systems*, *Computer Science Artificial Intelligence*, *Computer Science Interdisciplinary Applications*, *Ergonomics*, *Telecommunications*, *Robotics*, *Instruments Instrumentation*, *Computer Science Theory Methods*, *Computer Science Software Engineering*, *Information Science Library Science*, *Engineering Multidisciplinary*, *Multidisciplinary Sciences*, *Materials Science Multidisciplinary*, *Computer Science Hardware Architecture*, *Automation Control Systems*, *Imaging Science Photographic Technology*, *Engineering Industrial*, *Transportation Science Technology*, *Engineering Civil*, *Green Sustainable Science Technology*, *Transportation*, *Biotechnology Applied Microbiology*, *Mechanical Engineering*, *Nanoscience Nanotechnology*, *Materials Science Textiles*, *Mechanics.*

We then narrowed down papers by selecting original and open-access journal papers. The resulting 173 articles were examined by reviewing their titles and abstracts. Two authors reviewed them independently to identify studies focusing on AT development and application, including design, prototype, and user-test, and studies that performed data collection and data analysis were included in the search. Articles in which the two authors disagreed on inclusion/exclusion were discussed until a consensus was reached between them, resulting in 90 papers. Further, four works were excluded based on full-text screening, and a final pool of 86 research articles was determined for full literature review in this study.

### 3.2 Descriptive Analysis

Descriptive analysis was conducted on the 86 selected papers to describe AT development trends in the literature. The analysis counted the number of papers from the following three perspectives. First, the types of technology the papers developed were examined. An initial open coding performed by the second author elicited a total of 24 codes associated with developed technology in each article. Some of them were relatively similar (e.g., "web interface" and "GUI"), whereas others had precise and distinct meanings (e.g., navigation system). Then, second-order coding was performed by discussion among two authors to review the initial coding and title/abstract of the papers. This process elicited 14 codes used to categorize the papers in this study. Secondly, users' skills/capabilities, which the ATs aimed at supporting, were examined and classified into vision, hearing (hearing and speech communication), mobility, and cognition (cognition and learning skills), as discussed in Section 2. Thirdly, a categorization regarding mainstream technologies was made. Whereas ATs have been traditionally defined as devices specifically developed for people with disabilities, which implicitly means for a certain number of individuals, mainstream technologies are intended for more generalized use by a broader population, rather than for use entirely or primarily by people with disabilities [20].

### 4 Results and Discussion

The literature review provided insightful results that allowed for the creation of evidence-based findings on developments and applications of ATs. The constant growth of AT development was found predominantly in navigation systems (30 papers) such as a wearable navigation device [21], marker detection using machine learning techniques [22], and mobility aid systems (11 papers) such as a smart walker [23]. Other types of developed technologies that are worth mentioning are visual aid, hearing aid and audio, computer, and mobile app systems to improve accessibility.

Fig. 1 shows the evolution of different types of technology developed over time and reported in selected articles.

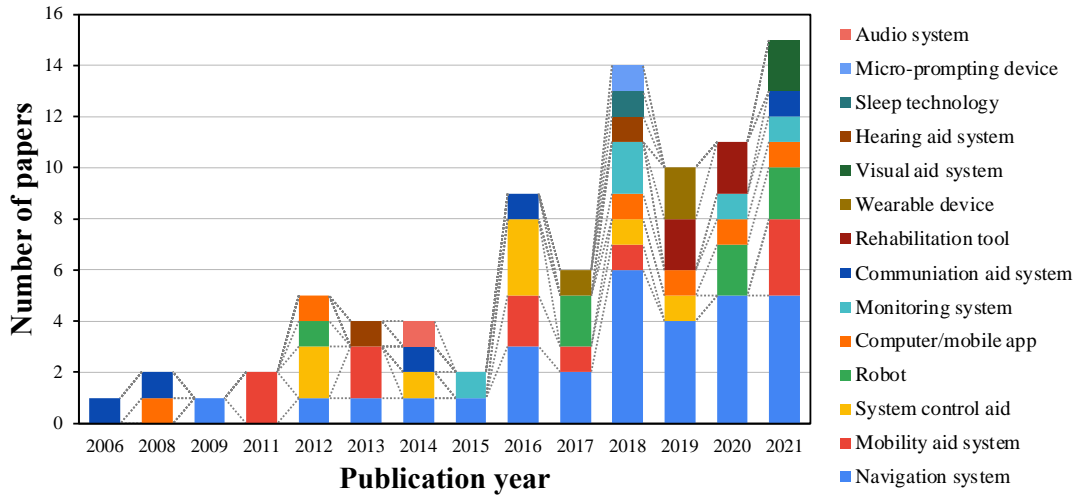

**Fig. 1.** Groups of type of technology studied and developed over the years.

Fig. 2 shows the variability of publications reporting AT developed to address different users' capabilities/skills. Around 39 papers focused on addressing visual capabilities, 26 mobility capabilities, 19 cognitive capabilities, and 10 hearing capabilities. The diagram shows a growth in the number of publications over the years, as well as a growth of studies for the development of AT to solve visual capability issues. Different patterns can be identified regarding the area of mobility or cognitive capabilities where the growth is not constant, as for visual capabilities. This trend may be attributed to more interest from the scientific community, more attention from a regulatory perspective, or more awareness of actual challenges to solve for the community.

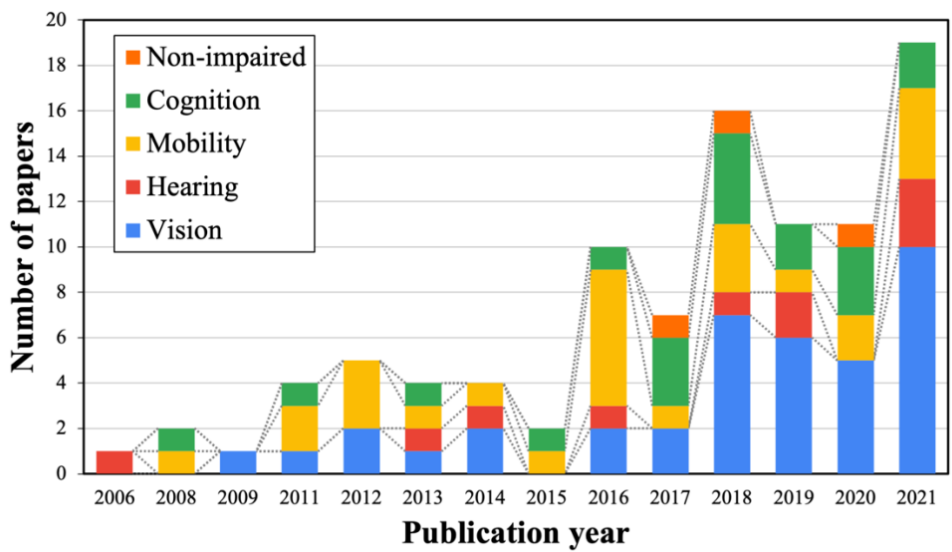

**Fig. 2.** Publications reporting AT developed to address different users' capabilities/skills.

Fig. 3 shows the trend of AT development in relation to mainstream or non-mainstream technologies. Mainstream technologies are those currently popular among a broader population and can be used, with some adaptations, to support people with different level of abilities. Non-mainstream refers to technologies developed specifically for a limited number of individuals with certain needs. Although the percentage of articles that reported research projects utilizing non-mainstream technology was higher than those utilizing mainstream technology, growth of the use of mainstream technology appears in recent years. Over the years, particularly from 2016, the number of products developed and tested in the selected papers based on mainstream technologies is growing. This dataset highlights a trend that will potentially impact the development of future AT and will lead to allowing AT to become more accepted and used by mainstream consumers.

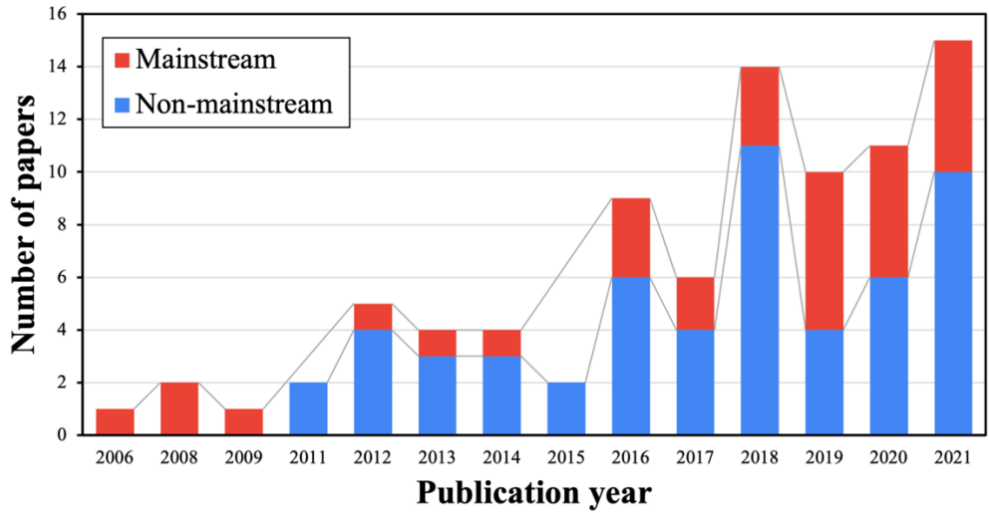

**Fig. 3.** Evolution of the trend of AT developed by using mainstream technologies (red) and non-mainstream technologies (blue).

Based on the results of the literature review, there is a strong focus on technologies developed to answer visual, hearing, and mobility challenges, which can vary from a severe impairment to a less severe or mild impairment. Notwithstanding the high number of technologies conceptualized and developed in the analyzed papers belonging to the category of AT, growth in the use of mainstream technologies can be highlighted as enabling the development of new AT. On the other side, the influence that AT has on mainstream technology appears evident. Some of them became widely used by mainstream users. This mutual effect might bring added value when developing technologies for a diverse range of people and the use of mainstream technologies can influence the development of AT and vice versa.

To reinforce this initial evidence with examples from daily technologies, three supporting devices that help to achieve certain tasks today were developed years ago as AT. Closed captioning service is a clear example of a digital technology that was born as an AT, to provide support for people who are deaf or hard of hearing [24].

Captions became particularly beneficial for people watching videos in their non-native language, for children and adults learning to read, and in general to facilitate the understanding of a topic in a video. This can be considered as an example of successful AT that became a mainstream technology appreciated by most users.

Assistive listening devices refer to various types of amplification equipment designed to improve the communication of individuals who are hard of hearing to enhance the accessibility to speech signal when individual hearing instruments are not sufficient [25]. Among them, noise-canceling headphones are these days among the most popular offerings of technology manufacturers. Once again, early AT developments turned out to become technologies that offer a variety of augmented capabilities to mainstream users.

Speech recognition systems and devices have proved to be enormously beneficial for people with physical disabilities, having the potential to provide a fast and easy-to-use means of input for computer access and control of the home environment [26].

These days smartphones, tablets, and computers have such capability and billions of users around the world are making use of the potential of the speech recognition systems. This is a further example of how an AT became more used as a mainstream technology.

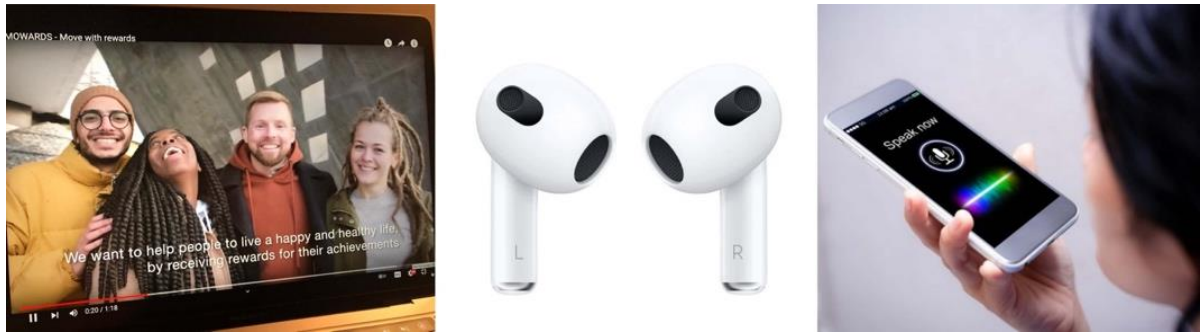

**Fig. 4.** Examples of early developments of Assistive Technology that has now become mainstream.

## 5    Conclusions and future vision

In this paper, we completed a systematic literature review that identified a growing trend of the evolution of certain AT as mainstream technologies.

It was identified that some of the major applications of these developments are having a significant impact on human capabilities such as mobility, hearing, and vision.

While AT has been defined as a term over the years that specifically focuses on products developed for people with special needs, we argue that such products may have a broader use across mainstream users. This research also provides evidence of the value of Inclusive Design (ID) in AT development. ID refers to a design process in which a mainstream product, service, or environment is designed to be usable by as many people as reasonably possible. AT design is a clear example of an approach that allows for the design of products for users with specific needs, which can then be easily extended to meet the needs of a broader population.

The examples mentioned in this paper clarify with evidence the relevance of designing with a specific goal in mind, that is, not to limit the creative process to solve a limited number of needs identified across a sample population that doesn't represent the variety of people, their capabilities, age, gender, language, and culture, but rather to go beyond the mainstream concept of designing for the "many" and start to include in the design process diverse groups of people [27].

This process will lead to the creation of products that are not only classified as AT but can serve a similar purpose of enabling all people, regardless of their age, gender, ability, and culture to be equal participants in the consumer world.

**Acknowledgments.** This project received funding from Horizon 2020: EU Programme for Research and Innovation under the Marie Skłodowska-Curie grant agreement No. 846284, and was partly supported by the Center of Innovation Program of the Japan Science and Technology Agency (Grant No. JPMJCE1309). The authors equally contributed to the conception, data analysis, development, and writing of this research article.